\documentclass[prl,amsmath,amssymb,floatfix]{revtex4}
\usepackage{natbib}
\usepackage{dcolumn}
\usepackage{graphicx}
\usepackage{bm}
\usepackage{scrtime}[24hr]
\usepackage[usenames]{color}

\begin{document}

\newcommand{\degc}{\ensuremath{^{\circ}}C}

\bibliographystyle{apsrev}

\title{Reply to comment on `Measurement of Effective Temperatures in an Aging Colloidal Glass'}
\author{Nils Greinert}
\author{Tiffany Wood}
\author{Paul Bartlett}

\affiliation{School of Chemistry, University of Bristol, Bristol BS8
1TS, UK.}



\maketitle

In a comment, Jop \textit{et al.} \cite{4901} have criticized the recent report \cite{4603} of an elevated effective temperature ($T_{\textrm{eff}}$) in an aging colloidal Laponite glass. Following a procedure similar to that used in \cite{4603}, Jop \textit{et al.} measured the effective temperature of a colloidal glass. They found that while $T_{\textrm{eff}}$ was constant at the beginning of the experiment, the measured temperatures became increasingly erratic when jamming occurred. As a consequence of the large statistical errors in their values of $T_{\textrm{eff}}$, the authors conclude that the results reported in \cite{4603} are inconclusive and that furthermore there is no increase in $T_{\textrm{eff}}$ with age in Laponite. The authors proposed a number of experimental artifacts which could account for this disagreement.

\begin{figure}[h]
\includegraphics[width=3.0in]{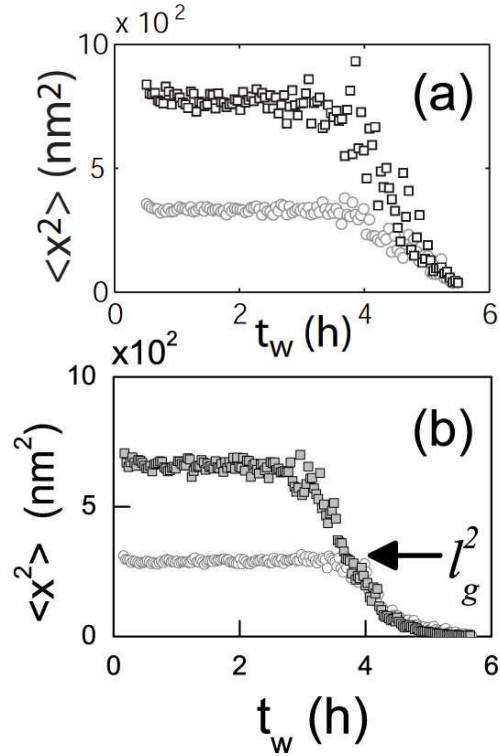}
\caption{The evolution of $\left <  x^{2} \right >$ with $t_{w}$. Comparison of different experiments. (a) Data from Fig.~1 of Ref.~\cite{4901}, 2.3 wt\% Laponite. Circles and squares denote $\left < x^{2} \right >$ measured for optical trap stiffnesses of $k_{2} = 14.4$ pN/$\mu$m and $k_{1} = 6.34$ pN/$\mu$m respectively. (b) Data from Fig.~2b of Ref.~\cite{4603}, 2.4 wt\% Laponite. Circles and squares correspond to $k_{2} = 11.0$ pN/$\mu$m and $k_{1} = 4.4$ pN/$\mu$m respectively.} \label{fig:msd}
\end{figure}

\begin{figure}[h]
\center{
\includegraphics[width=4.0in]{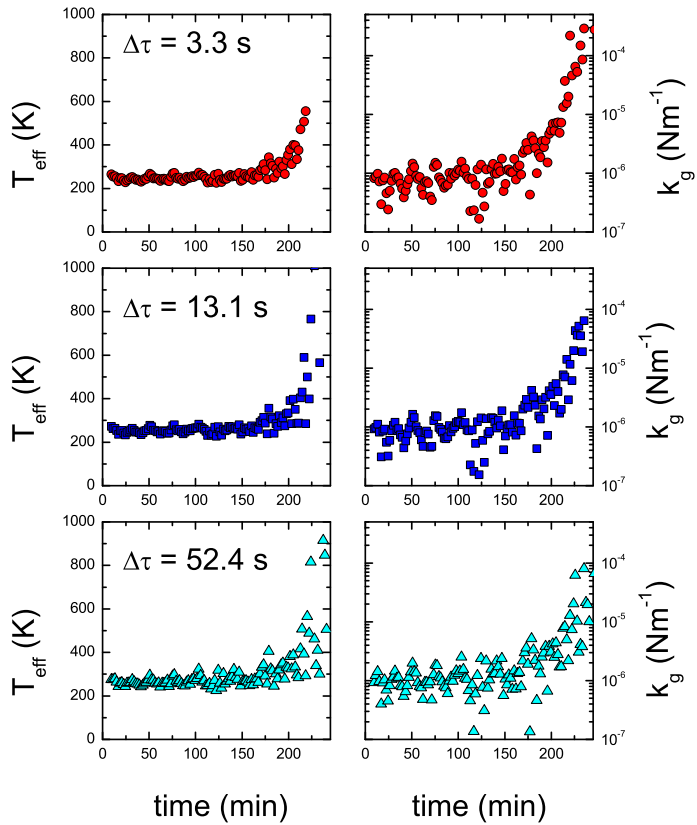}
\caption{(color online). Evolution of the effective temperature ($T_{\textrm{eff}}$) and glass elasticity ($k_{g}$) with aging time. Conditions as \cite{4603}. The mean squared displacement $\left < x^{2} \right >_{\Delta \tau}$ is determined using a time window $\Delta \tau$: from top to bottom $\Delta \tau = 3.3$ (red $\circ$), 13.1 (blue $\square$) and 52.4 s (cyan $\vartriangle$).}
 \label{fig:time-windows}
 }
\end{figure}

We make three remarks. First, the experiments of Jop \textit{et al.} are significantly noisier than those reported in \cite{4603}. In consequence the effective temperatures computed in \cite{4901} have \textit{larger} uncertainties than the data of Ref.~\cite{4603}. This is readily seen from a visual comparison of the reported mean squared displacements $\left <  x^{2} \right >$. Measurements made under comparable conditions are reproduced in Fig.~\ref{fig:msd}. The variability of the $\left <  x^{2} \right >$ values (plotted in Fig.~\ref{fig:msd}a) is clearly larger than the data measured under similar conditions and reported in Ref.~\cite{4603} (reproduced in Fig.~\ref{fig:msd}b). The comparison is particularly pronounced in the vicinity of the jamming transition ($t_{w} \sim 3.5$ h) where the fluctuations of the probe particle are restricted first by the aging glass. High quality data in this region is key to observing the increase of $T_{\textrm{eff}}$ with age. The large scatter in $\left <  x^{2} \right >$ for $t_{w} \sim 3.5$ h evident in Fig.~\ref{fig:msd}a explains the rather large measuring errors seen in the temperatures reported in \cite{4901}. The effect of these errors is to obscure any underlying temperature trend. In contrast, the more reproducible data of \cite{4603} reveals that $T_{\textrm{eff}}$ actually increases with $t_{w}$. Interestingly, we note from Fig.~\ref{fig:msd} that the shape of the data sets from Ref.~\cite{4901} and \cite{4603} are  remarkably similar, which suggests that reducing the uncertainty in the experiments of Ref.~\cite{4901} would reveal a very similar age-dependent $T_{\textrm{eff}}$ to that reported in \cite{4603}. It is not clear what are the causes of the large scatter seen in the data of Jop \textit{et al.} but Laponite samples are notoriously difficult to handle. We found that  a rigorous sample preparation regime was essential to achieve reproducible data \cite{1}.

Second, Jop \textit{et al.} suggest that the duration $\Delta \tau$ of the time window used to determine $\left <  x^{2} \right >$  may be too small to include all long lived fluctuations in the aging glass. We have checked for this eventuality by increasing $\Delta \tau$ and re-analysing the data presented in \cite{4603}. The results are shown in Fig.~\ref{fig:time-windows}.   Clearly while broadening the time window increases the experimental scatter, the primary conclusion of Ref.~\cite{4603} remains unaltered with an increase in the effective temperature evident at large $t_{w}$.

\begin{figure}[h]
\includegraphics[width=6.0in]{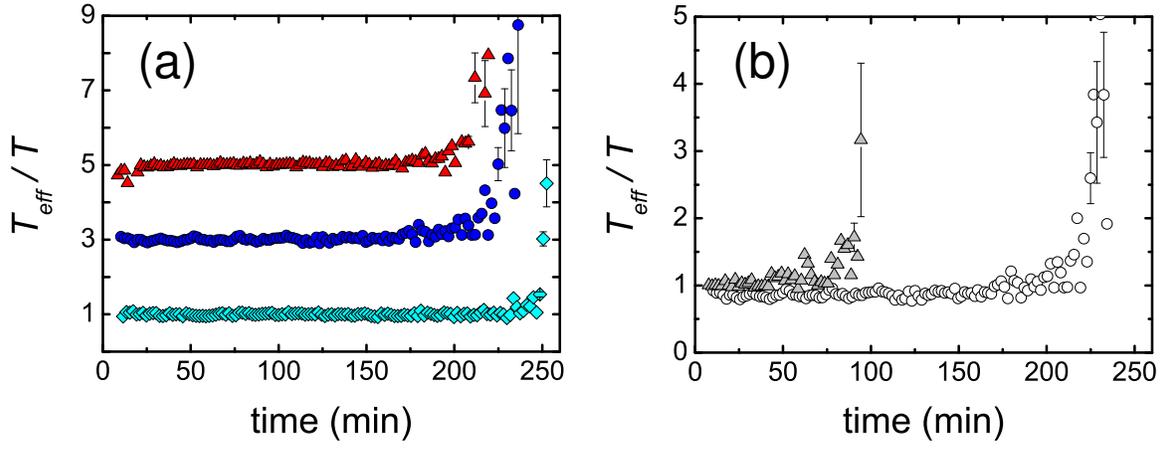}
\caption{(color online). Evolution of the scaled effective temperature $(T_{\textrm{eff}} / T)$  with $t_{w}$. $T$ is the room temperature. (a) Repeat samples of 2.4 wt\% Laponite. Curves have been vertically displaced by two units for clarity. (b) Effect of Laponite concentration: $\circ$ 2.4, and $\vartriangle$ 2.9 wt \%.}
\label{fig:repeat}
\end{figure}

Finally, Jop \textit{et al.} report difficulties observing any increase in the effective temperature despite using a variety of different experimental conditions. Again, this is not our experience.
We have repeated the measurements outlined in Ref.~\cite{4603} several times and Fig~\ref{fig:repeat}(a) illustrates typical results. The effective temperatures determined from three physically-distinct suspensions containing 2.4 wt\% Laponite are shown (neighboring plots are displaced two units vertically for clarity). Although there are small shifts between experiments (particularly along the time axis) the reproducibility is generally high. Note in each case an increase in the effective temperature is seen with $t_{w}$. Similar results are obtained at different Laponite concentrations (Fig~\ref{fig:repeat}(b)).

In conclusion, our results confirm the validity of the experiments outlined in Ref.~\cite{4603} and show that the effective temperature of a glass increases with age. The artifacts pointed out by Jop \textit{et al.} \cite{4901} do not alter significantly this conclusion.




\end{document}